\documentstyle[aps,prb,amssymb]{revtex} 
\input epsf.sty 

\begin{document}
\setcounter{figure}{0} 
\renewcommand{\theequation}{\arabic{equation}} 

\twocolumn[ 
\hsize\textwidth\columnwidth\hsize\csname@twocolumnfalse\endcsname 
 
\draft 

\title{Deconfinement Phase Transition in a 3D Nonlocal U(1) 
Lattice Gauge Theory} 
\author{Gaku Arakawa and Ikuo Ichinose } 
\address{Department of Applied Physics, 
Nagoya Institute of Technology, 
Nagoya, 466-8555 Japan } 
\author{Tetsuo Matsui} 
\address{Department of Physics, Kinki University, 
Higashi-Osaka, 577-8502 Japan } 
\author{Kazuhiko Sakakibara}
\address{
Department of Physics, Nara National College of Technology, 
Yamatokohriyama, 639-1080 Japan
} 
\date{\today}  
 
\maketitle

\begin{abstract}   
We introduce a 3D compact
U(1) lattice gauge theory having
 nonlocal interactions in the temporal direction, and 
study its phase structure.
The model is relevant for the compact QED$_3$ and strongly correlated 
electron systems like the t-J model of cuprates.
For a power-law decaying long-range interaction, 
which simulates the effect of gapless matter fields, 
a second-order phase transition takes place separating the 
confinement and deconfinement phases. For 
an exponentially decaying inter- 
action simulating matter fields with gaps, 
the system exhibits no signals of a second-order transition. 

\vspace{0.3cm}
DOI: 10.1103/PhysRevLett.94.211601 \hspace{2cm} PACS numbers:\ 11.15.Ha, 12.38.Gc, 71.27.+a\\
\end{abstract} 

]

 
\setcounter{footnote}{0}
\setcounter{equation}{0} 

The U(1) lattice gauge theory (LGT) in three dimensions (3D) 
coupled to matter fields describes various interesting physical systems.
The compact QED$_3$ is just a such system and its phase
structure has been studied by various methods\cite{QED}.
In condensed matter physics,
interesting observations were made that 
the strongly-correlated electron systems in two dimensions 
like
the antiferromagnetic Heisenberg spin model, 
the t-J model of high-$T_c$ cuprates, and 
the fractional quantum Hall states  are described 
by 3D U(1) gauge theories due to the introduction
of auxiliary collective fields\cite{im1,heisenberg,yoshioka,FQH}.

For the case {\em without matter-field couplings},
Polyakov \cite{polyakov} showed that the 3D compact U(1) gauge theory
is always in the confinement phase due to the 
monopole (instanton) condensation. 
For the case with {\em couplings to matter fields}, there is still no 
consensus on the question of whether the system in two spatial dimensions 
exhibits a phase transition into a deconfinement phase.
Probably the answer depends on the properties of coupled matter fields.  
This question is important because  the 
``fractionalization" of electrons may be interpreted as 
the deconfinement phenomenon of U(1) gauge dynamics.
For the t-J model, the possibility of charge-spin separation
(CSS) is of great interest since it may explain various
anormalous behaviors of the metallic state of 
cuprates\cite{anderson}.
In Ref.\cite{css} it was argued that the CSS takes place
{\it below} certain critical temperature $(T)$ as a 
deconfinement (perturbative) phase of
an effective U(1) LGT which is derived by the hopping expansion
of spinons and holons in the slave-particle representation
at finite $T$. In related works, Nagaosa\cite{nagaosa} argued 
that the deconfinement phase
  is possible {\it above} some $T$, whereas Nayak\cite{nayak} argued
that deconfinement does not occur at any $T$. 
The deconfinement
phase at $T=0$ for systems with gapless excitations 
are supported in Ref.\cite{css+} and denied in Ref.\cite{css-}. 

In this Letter, we introduce and study 
a LGT with {\it nonlocal} interactions in order
to investigate the phase structure of compact U(1) gauge theories 
coupled to matter fields on the  3D lattice (2D system at $T=0$).  
We first consider the cases of massless and massive  
{\it relativisitic} matter fields. Then we apply the model
to the nonrelativistic electron systems.  The results of this paper for 
gapless excitations in electron systems shall
complement our previous result of CSS\cite{css} because  
the hopping expansion employed there 
may be inadequate for massless excitations at $T=0$.  

Let us start with the path-integral representation of the partition 
funciton $Z$ of gauge field $U_{x\mu}$ and (bosonic and/or
fermionic) matter fields $\phi_x$,
\begin{eqnarray}
Z &=& \int\prod_{x}d\bar{\phi}_x d\phi_x \prod_{x,\mu}dU_{x\mu}
\exp(A), \nonumber\\
A &=& -\sum_{x,y}\bar{\phi}_x \Gamma_{xy}(U) \phi_y
+A_U, \nonumber\\
A_U&=& q\sum_{x, \mu < \nu}(\bar{U}_{x\nu}
\bar{U}_{x+\hat{\nu},\mu}U_{x+\hat{\mu},\nu}U_{x\mu}+ c.c.),
\label{original}
\end{eqnarray}
where $x=(x_0,x_1,x_2)$ 
is the site-index of the 3D lattice of the size 
$V = N_0 N_1 N_2$ with the periodic boundary condition, 
$\mu (=0,1,2)$ is the (imaginary) time and spatial
direction index, 
$\phi_x$ is the matter field on $x$,
$U_{x\mu}=\exp(i\theta_{x\mu}) \; (-\pi < \theta_{x\mu} 
\le \pi)$ is the $U(1)$ gauge variable
on the link $(x,x+\hat{\mu})$,
$\Gamma_{xy}(U)$ represents the {\it local}
couplings of $\phi_x$ to $U_{x\mu}$.

By integrating over $\phi_x$, 
we obtain an effective gauge theory,
\begin{eqnarray}
Z &=& \int\prod_{x,\mu}dU_{x\mu}
\exp\Big[
f\; {\rm Tr}\;  \log\; \Gamma_{xy}(U) +A_U\Big],
\label{effective-model}
\end{eqnarray}
where $f$ is a parameter counting 
the statistics and internal degrees of freedom of $\phi_x$.
Due to the $({\rm Tr}\;  \log)$ term, the effective gauge theory becomes 
{\it nonlocal}. 
For relativistic matter fields, it is expanded as a sum over
all the closed random walks ${\cal R}$ (loops including
backtrackings) on the 3D lattice which  represent world lines 
of particles and antiparticles as
\begin{equation}
{\rm Tr}\;  \log\; \Gamma_{xy}(U)= \sum_{\cal R}
\frac{\gamma^{L[{\cal R}]}}{L[{\cal R}]}
\prod_{(x\mu) \in {\cal R}}U_{x\mu}.
\label{eff1}
\end{equation}
$L[{\cal R}]$ is the length 
of ${\cal R}$, and $\gamma =(6+m^2)^{-1}$ is the hopping parameter
(6 is the number of links emanating from each site, and
$m$ is the mass
of the matter field in unit of the lattice spacing).  
For the constant gauge-field configuration $U_{x\mu}=1$, the expansion
in (\ref{eff1}) is logarithmically divergent $\sim\log m$
as $m \rightarrow 0$
due to the lowest-energy zero-momentum mode. 

Below we shall study a slightly more tractable model than  
that given by Eq.(\ref{effective-model}). It is suggested from
the formal hopping expansion (\ref{eff1}),
and  obtained by retaining only the rectangular loops  
extending in the temporal direction in the loop sum 
and choosing their coefficients optimally as follows;
\begin{eqnarray}
Z_{\cal T} &=& \int\prod_{x,\mu}dU_{x\mu}
\exp(A_{\cal T}), \nonumber\\
A_{\cal T} &=& g\sum_{x}\sum_{i=1}^2
\sum_{\tau=1}^{N_0} c_{\tau}(V_{x,i,\tau}
+\bar{V}_{x,i,\tau})
+A_S,\nonumber\\
V_{x,i,\tau} &=& \bar{U}_{x+\tau\hat{0},i}
\prod_{k=0}^{\tau-1}[\bar{U}_{x+k\hat{0},0}
U_{x+\hat{i}+k\hat{0},0}]U_{xi},\nonumber\\
A_S &=& g \lambda\sum_x(\bar{U}_{x2}
\bar{U}_{x+\hat{2},1}U_{x+\hat{1},2}U_{x1}+ c.c.).
\label{simplified-model}
\end{eqnarray}
$V_{x,i,\tau} $ 
is the product of $U_{x\mu}$ along the rectangular
$(x, x+\hat{i}, x+\hat{i}+\tau \hat{0},x+ \tau \hat{0})$ 
of size $(1 \times \tau)$ in the $(i-0)$ plane. 
In $A_S$, we have retained only the nearest-neighbor spatial coupling.
For the nonlocal coupling constant $c_{\tau}$,
we consider the following two cases; 
\begin{eqnarray}
c_{\tau} &=& \left\{
\begin{array}{ll}
 \frac{1}{\tau}, & {\rm power-law\ decay\ (PD)}, \\
e^{-m \tau}, & {\rm exponential\ decay\  (ED)}. \\
\end{array}
\right.
\label{ctau}
\end{eqnarray}
The power $-1$ in the PD case in (\ref{ctau})
reflects the effect of massless excitations. 
In fact, this $c_{\tau}$ generates 
a logarithmically divergent action for $U_{x\mu}=1$
explained below Eq.(\ref{eff1}) 
 as one can see from the relation,
$\sum_\tau \exp(-m \tau)\tau^{-1} \simeq \log (1/m)$.
The action for $m=0$ is then proportional to
$\sum_\tau \tau^{-1} \simeq \log N_0$ for finite $N_0$. 
On the other hand, the ED model contains $m$ and simulates the 
case of massive matter fields\cite{ed}.

We made Monte Carlo
simulations to determine the phase structure of this model. 
We consider the isotoropic lattice,
$N_\mu = N$,  with the periodic boundary condition up to
$N = 32$, where the limit $N\rightarrow  \infty$ corresponds to the
system on a 2D spatial lattice at $T=0$.
For the mass of the ED model,
we set $m=1$. 
For the spatial coupling $\lambda$
scaled by $g$, we consider the two typical cases
$\lambda=0$ (i.e., no spatial coupling) and $\lambda=1$.

First, we calculate the following average $E$ (``internal energy")
and the fluctuation $C$ (``specific heat") of $A_{\cal T}$;


\begin{figure}[thbp]
\begin{center}
\epsfxsize=8.5cm
\hspace{-0.3cm}   
\epsffile{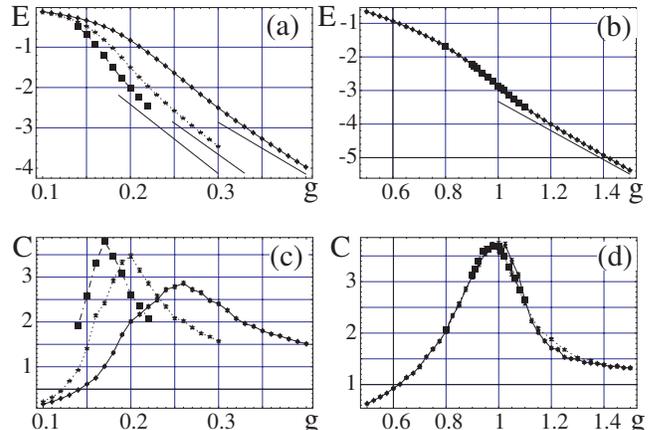}
  \caption{Internal energy $E$ and the fluctuation $C$ of the action  
with $\lambda=1$ vs non-local coupling $g$ 
for $N =8 (\blacklozenge), 16(\bigstar), 24(\blacksquare)$;
(a,c) PD model, (b,d) ED model.
The solid lines in (a) and (b) are the large-$g$ expansion.
In the PD model, strong $N$ dependence is observed in $E$ 
at large $g$ and in the developing peak of $C$. 
They indicate a second-order phase transition in the PD model.
}
\label{fig-1}
\end{center}
\end{figure}

\begin{equation}
E \equiv -\langle A_{\cal T}\rangle/V, \ \
C \equiv \langle (A_{\cal T}- \langle A_{\cal T}\rangle)^2\rangle/V.
\label{EC}
\end{equation}  
For small $g$, the high-temperature expansion (HTE) gives 
$Z \simeq\exp[g^2(2Q_2+\lambda^2)V]$
($Q_k\equiv\sum_\tau c_{\tau}^k$), 
whereas for large $g$, the low-temperature expansion (LTE)
around $V_{x,i,\tau} =1$ gives 
$Z \simeq \exp[(4gQ_1+ 2g \lambda-\log g)V]$.

In Fig.1, we present $E,C$ for $\lambda=1$ vs the
nonlocal coupling $g$. They agree with the above HTE and LTE. 
In the PD model, $E$ of Fig.1(a) connects the HTE result and 
LTE result, showing that $V_{x,i,\tau} \sim 1$
for large $g$.
$C$ of Fig.1(c) shows that its peak develops 
as $N$ increases.  
The finite-size scaling analysis shows $C$ of Fig.1(c) fits 
well in the form,
$C(g,N)=N^{\sigma/\nu}\phi(N^{1/\nu}\epsilon), 
\epsilon=(g-g_c)/g_c$ with $
\nu= 1.2\sim 1.3, \sigma/\nu= 0.25\sim0.26, g_c
= 0.10\sim 0.12.$ 
These results indicate that the PD model exhibits 
a second-order phase transition separating the disordered
(confinement) phase and the ordered (deconfinement) phase
at $g=g_c$.
  This transition will be confirmed later by the measurement of
Polyakov lines.
On the other hand, 
the peak of $C$ in the ED model does not develop as $N$ increases,
showing  {\em no} signals of a second-order 
transition. It may have a higher-order transition or just  
a {\em crossover}.
Simulations of the models with $\lambda=0$ 
give similar behaviors of $E,C$, preserving the above  phase structure
for $\lambda=1$.

To study the nature of gauge dynamics, 
we calculated the 
spatial correlations of Polyakov lines $P_{x_{\bot}}$;
\begin{eqnarray}
&& P_{x_{\bot}} = \prod_{x_0=1}^{N_0}U_{x_{\bot},x_0,0},\ 
(x_{\bot}=(x_1,x_2)), \nonumber \\
&& f_P(x_{\bot}) = \langle \bar{P}_{x_{\bot}} P_{0}\rangle.
\label{Plines}
\end{eqnarray}
Since the present model (\ref{simplified-model}) contains
no long-range interactions in the spatial directions, 
$f_P(x_{\bot})$ is expected 

\begin{figure}[tbp]
\begin{center}
\epsfxsize=8.5cm     
\epsffile{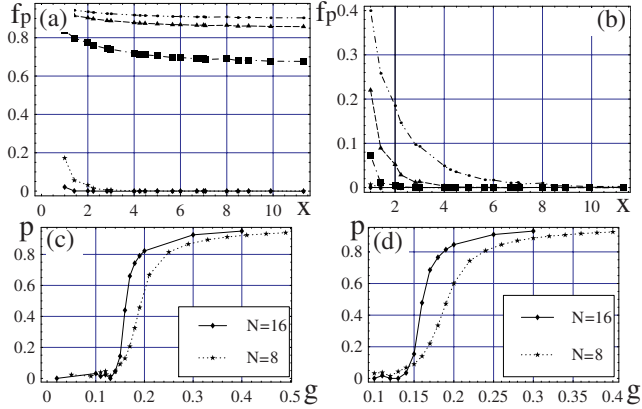}
\caption{Correlations of Polyakov lines, $f_P(x_{\bot})$,
vs $|x_{\bot}|$.
(a) PD ($\lambda=0$, N=16) with $g= 0.4,0.3,0.2,0.1,0.02$ from above.
(b) ED ($\lambda=0$, N=16) with $g= 2.5,2.0,1.5,1.0,0.5$ from above.
(c) and (d) show the order parameter $p=[f_P(x_{\bot}^{\rm MAX})]^{1/2}$ 
vs $g$ for the PD model; (c) $\lambda=0$ and  (d) $\lambda=1$. 
They  exhibit a long-range order for $g > g_c \simeq 0.15$
  in the PD model.
}
\label{fig-2}
\end{center}
\end{figure}

\noindent
to supply us with a good order parameter
to detect a possible confinement-deconfinement transition.
The deconfinement phase is characterized by small fluctuations 
of $U_{x\mu}$ and therefore by an  order in $f_P(x_{\bot})$.

In Fig.2, we present  $f_P(x_{\bot})$ with $\lambda=0$.
The PD model of Fig.2(a)  clearly exhibits an off-diagonal long-range order,
i.e., $\lim_{x_\bot \rightarrow \infty} f_P(x_{\bot}) \neq 0 $
for $g \geq 0.20$, whereas the ED model of Fig.2(b) {\it does not} for all 
$g$'s.
To see this explicitly, we plot in Fig.2(c,d) the order 
parameter $p 
\equiv [f_P(x_{\bot}^{\rm MAX})]^{1/2} $ for
the PD  model, 
where $x_{\bot}^{\rm MAX} \equiv N/\sqrt{2}$ is the 
distance at which $f_P$ becomes minimum 
due to the periodic boundary condition.
$p$ starts to develop continuously from zero  
at $ g = g_c \simeq 0.15$. The size dependence of $p$
shows a typical behavior of a second-order transition.
Thus  the gauge dynamics of the PD  model is 
realized in the deconfinement phase for $g > g_c$,
whereas it is in the confinement phase for $g < g_c$. 
In contrast, the ED model stays 
always in the confinement phase.
These results including the value of $g_c$ are consistent with 
those derived from the data of $E, C$ in Fig.1.

To see the details of gauge dynamics, we measured
the instanton density $\rho(x)$, an index for disorder
of $U_{x\mu}$. 
We employ the definition of $\rho(x)$ in U(1) LGT by
DeGrand and Toussaint \cite{instanton}. 
For the {\it local} 3D compact U(1) LGT without matter 
fields, the average density $\rho = \langle \sum_x|\rho(x)| \rangle/V$ 
was measured in Ref.\cite{instanton2}.
In Fig.3 we present 
$\rho$ vs $g$. It  decreases as $g$ increases  more rapidly in 
the PD  model than in the ED model. 
This behavior of $\rho$ is consistent with the result 
that the PD model exhibits a second-order transition, while
the ED model does not.
The $\lambda$ coupling
enhances the rate of decrease in $\rho$ 
as one expects
since the spatial coupling favors the ordered
deconfinement phase. 
In the ED model 
with $\lambda=1$, $\rho$ is fitted by $\propto \exp(-cg)$ 
with a constant $c$
in the dilute (large $g$) region, and the  smooth increase for smaller $g$

\begin{figure}[thbp]
\begin{center}
\epsfxsize=7cm     
\epsffile{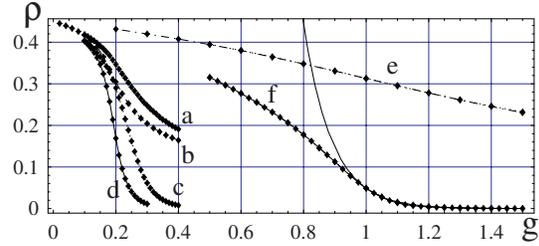}
\caption{
Average instanton density $\rho$ vs $g$.
(a)PD($\lambda=0$, $N=8$), (b)PD($\lambda=0$, $N=16$),
(c)PD($\lambda=1$, $N=8$), (d)PD($\lambda=1$, $N=16$),
(e)ED($\lambda=0$, $N=8,16$),
(f)ED($\lambda=1$, $N=8,16$).
The solid curve $\propto \exp(-cg)$ fits (f) at large $g$. 
}
\label{fig-3}
\end{center}
\end{figure}

\vspace{-0.5cm}
\begin{figure}[htbp]
\begin{center}
\epsfxsize=8cm     
\epsffile{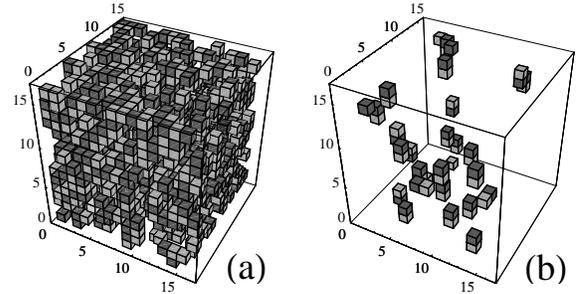}
\caption{
Snapshots of instanton configuration $\rho(x)$ for N=16. 
(a) PD($\lambda=1, g= 0.15$) and (b) PD($\lambda=1, g= 0.30$).
The light cubes for $\rho(x)=1$ and the dark cubes 
for $\rho(x)=-1$. }
\label{fig-4}
\end{center}
\end{figure}

\noindent
indicates  a crossover from 
the dilute gas of instantons 
to the dense gas, just the  behavior similar to the case of
pure and local LGT\cite{polyakov,instanton2}. 
 
 In Fig.4 we show snapshots of $\rho(x)$ 
for the PD model with $\lambda=1$. Fig.4(a) is a dense gas
and Fig.4(b) is a dilute gas. They are separated at 
$g_c \simeq 0.20$, the location of the 
peak of $C$ for $N=16$.
In Fig.4(b), instantons mostly appear in dipole pairs 
at nearest-neighbor sites, $\rho(x)=1, \rho(x\pm \mu)=-1$,
 while in Fig.4(a), they appear densely and it is 
 hard to determine their partners.
In both cases, the distributions $\rho(x)$ have no apparent
 anisotoropies like column structures. However, the orientations
of dipoles in Fig.4(b) are mostly ($\sim 92\%$) in the temporal direction 
as expected from Eq.(\ref{simplified-model}).
  
For ordinary {\it pure and local} gauge systems,  Wilson loop 
$W[{\cal C}]\equiv \langle \prod_{\cal C} U_{x\mu} \rangle$ along
a closed loop ${\cal C}$ on the lattice is used 
as  an order parameter to study the gauge dynamics;
$W[C]$ obeys the area law in the confinement phase and
the perimeter law in the deconfinement phase; 
\begin{eqnarray}
W[{\cal C}] \sim
\left\{
\begin{array}{ll}
\exp(-a S[{\cal C}]),\ & {\rm area\ law}, \\ 
\exp(-a' L[{\cal C}]),\ & {\rm perimeter\ law},
\end{array}
\right.
\label{wilson-loop}
\end{eqnarray}
where $S[{\cal C}]$ is the minimum area of a surface, the boundary of 
which is ${\cal C}$.
For a  local LGT  containing matter fields, $W[{\cal C}]$ cannot be  
an order parameter because the 
matter fields generate the terms
$\prod_{\cal C} U_{x\mu}$ 
with coeffcients $\sim \exp(-b L[{\cal C}])$ in the effective action. 
However, in the

\begin{figure}[thbp]
\begin{center}
\epsfxsize=8.5cm    
\epsffile{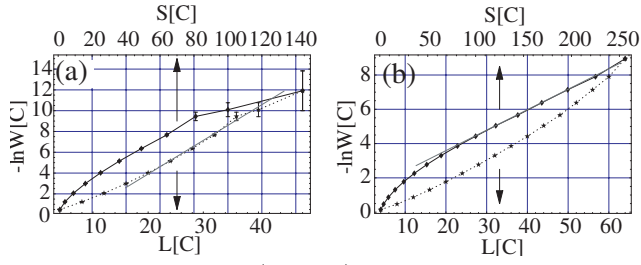}
\caption{Wilson loops ($N=32$) in the 1-2 plane
at large $g$  vs $L[\cal C]$ or $S[\cal C]$. 
(a) PD($\lambda=1,g =0.25$), (b) ED($\lambda=1,g= 1.5$). 
The PD model seems to prefer the perimeter law,
whereas the ED model prefers the area law.}
\label{fig-5}
\end{center}
\end{figure}

\noindent
present model (\ref{simplified-model}), 
the nonlocal terms are restricted only along
the temporal direction, so it is interesting to measure
$W[\cal C]$ for the loops 
lying in the {\it spatial (1-2) plane}.

In Fig.5, we plot $W[{\cal C}]$. 
For the PD model in Fig.5(a), the data at $g=0.25$ seem to prefer
 the perimeter law. For the ED model in Fig.5(b), 
the area law fits $W[{\cal C}]$  
better than the perimeter law at $g = 1.5$;
a considerably larger value than $g \simeq 1.0$ at the peak
of $C$. This suggests the area law holds in the ED model
at all $g$. These observations are consistent with the previous results
on the (non)existence of a  phase transition.
Wilson loops in the spatial plane
are useful to study the gauge dynamics of the present model.

We have observed that the
nonlocal couplings along the temporal direction in the PD model 
have sufficient effect of suppressing  fluctuations of $U_{x\mu}$ 
to produce the deconfinement phase. This result strongly suggests
a deconfinement transition in the original model (\ref{effective-model})
with massless matter fields,
because the  isotropically distributed nonlocal couplings of the original 
model should have similar effect.
In such a possible deconfinement phase of the original model, 
perturbation theory may be applicable, which 
predicts the potential energy
 between two charges as  $V(r) \sim r^{-1}$, a weaker one than
the 3D Coulomb potential $V(r) \sim \log(r)$.

Let us turn to the nonrelativistic case.
For the t-J model, by using the hopping expansion of holons and spinons 
at finite $T$ with the continuous imaginary time, 
we derived an effective gauge
theory, which is highly nonlocal in the temporal direction.
The obtained effective theory has a similar action
as Eq.(\ref{simplified-model}) with $c_\tau=$constant and 
$g \propto n$ where $n$ is the density of matter 
fields(holons and spinons)\cite{css}.
The nonlocal correlations of the effective gauge field like 
$c_\tau=$constant
come from the fact that the nonrelativistic fermions contain a higher 
density of low-lying excitations compared to Dirac fermions, i.e., the 
existence of the Fermi surface (or line).
Although the above effective gauge model is obtained for the system 
at finite $T$, we expect that a similar gauge model appears as an 
effective model at $T=0$.
Then it is interesting and also important to investigate the gauge model
(\ref{simplified-model}) with $c_\tau=$constant, the nondecaying (ND) model.
The ND model should shed some light on the anormalous normal state 
of cuprates because the Fermi line exists there.

We also made a Monte Caro simulation of  the ND model, 
and obtained a phase transition similar to the PD model.
However, the developing peak of $C$ shifts 
to smaller $g$ as $N$ increases more quickly  
than the PD model as  $g_c \sim 0.1 (N=8), 
0.045 (N=16), 0.03 (N=24)$ for $\lambda=1$.
It seems likely that $g_c \rightarrow 0$ as $N \rightarrow \infty$,
that is, the deconfinement phase dominates for all $g (>0)$.
This may be related with the diverging coefficient $Q_2(\propto N)$ 
in HTE, which
implies that the radius of convergence is zero.
This dominance of deconfienment phase of the ND model at $T=0$
may support the  CSS  at finite $T$, which is  
consistent with the result of Ref.\cite{css}.  
In contrast to the ND model, the PD model has a finite limit of 
$Q_2(=1.645)$, 
and has a finite region $0 \leq g \leq g_c$ of the confinement phase.




\begin{references}
\bibitem{ed}
For the ED case, we used Eq.(\ref{ctau}) instead of
$\exp(-m\tau)/\tau$ to make the comparison with the PD case more definitive.
\end{references}
\end{document}